\documentclass[aps,twocolumn,floats,prd,nofootinbib,10pt,longbibliography,superscriptaddress]{revtex4-1}

\usepackage{comment}
\usepackage[dvips]{graphicx} % 
\usepackage{graphicx,amsmath,amsfonts,amssymb,slashed,float,hyperref}
\usepackage[normalem]{ulem}
\usepackage{bbold,wasysym}
\usepackage{graphicx}
\usepackage{array,multirow}
\usepackage[utf8]{inputenc}
\usepackage{scalerel}
\usepackage{cleveref}

\usepackage[usenames,dvipsnames]{xcolor} 

\usepackage{soul}
\usepackage{bm}

\definecolor{RoyalBlue}{rgb}{0.25,.41,.88}
\definecolor{celestialblue}{rgb}{0.29, 0.59, 0.82}

\setstcolor{Blue}

\def\AB#1{\textcolor{Magenta}{AB}}

\newcommand{\be}{\begin{equation}}
\newcommand{\ee}{\end{equation}}
\newcommand{\bea}{\begin{eqnarray}}
\newcommand{\eea}{\end{eqnarray}}
\newcommand{\Beq}{\begin{equation}\begin{aligned}}
\newcommand{\Eeq}{\end{aligned}\end{equation}}

\definecolor{cerulean}{rgb}{0., 0.62,0.7}

\newcommand{\Mpl}{M_{\rm pl}}

\usepackage{color}
\usepackage{ifthen}
\newboolean{editorial}
\setboolean{editorial}{true}
\newcommand{\editorial}[2]{\ifthenelse{\boolean{editorial}}{\textcolor{red}{[\textsf{\textbf{{#1}}}: }\textcolor{blue}{\textsf{{#2}}}\textcolor{red}{]}}{}}

\usepackage{xcolor}

\begin{document}

\title{Gravitational Waves from Kinetic Preheating}

\author{Peter Adshead}
\affiliation{Illinois Center for Advanced Studies of the Universe \& Department of Physics, University of Illinois at Urbana-Champaign, Urbana, IL 61801, U.S.A.}

\author{John T. Giblin, Jr.}
\affiliation{Department of Physics, Kenyon College, Gambier, Ohio 43022, U.S.A.}
\affiliation{CERCA/ISO and Department of Physics, Case Western Reserve University, Cleveland, Ohio 44106, U.S.A.}
\affiliation{Center for Cosmology and AstroParticle Physics (CCAPP) and Department of Physics, The Ohio State University, Columbus, OH 43210, USA}

\author{Avery Tishue}
\affiliation{Illinois Center for Advanced Studies of the Universe \& Department of Physics, University of Illinois at Urbana-Champaign, Urbana, IL 61801, U.S.A.}

%\date{September 2023}
\begin{abstract}
We study gravitational wave production during kinetic preheating after inflation with a focus on scenarios that arise in $\alpha$-attractor models where a scalar dilaton-like inflaton is kinetically coupled to a second scalar field. We present high-resolution lattice simulations of three $\alpha$-attractor models for a range of parameters to probe regions where preheating is efficient.  We find that preheating in these models can be extremely violent, resulting in gravitational wave energy densities that can be constrained by cosmic microwave background measurements of the effective number of relativistic species, $N_{\rm eff}$. 
\end{abstract}

\maketitle

\section{Introduction \label{sec:intro}}
An early phase of accelerated expansion, inflation \cite{Guth:1980zm, Starobinsky:1980te,  Linde:1981mu,Albrecht:1982wi, Linde:1983gd},  solves the horizon and flatness problems of the hot Big Bang cosmology. Quantum vacuum fluctuations of the fields and metric during inflation are stretched outside the horizon before later reentering to seed the density fluctuations that eventually  give rise to the inhomogeneities of the cosmic microwave background (CMB) and the large scale structures in the Universe today \cite{Mukhanov:1981xt, Guth:1982ec, Hawking:1982cz, Bardeen:1983qw}. However, the microphysical origin of the accelerated expansion is far from understood. The simplest models for inflation (for example, Refs.\ \cite{Linde:1983gd, Freese:1990rb}) are now strongly disfavored \cite{Planck:2018jri, BICEP:2021xfz} and non-minimal inflationary mechanisms (for example, Starobinsky inflation \cite{Starobinsky:1980te}, Higgs inflation \cite{Bezrukov:2007ep}, and $\alpha$-attractors \cite{Kallosh:2013maa}) are now the leading candidates for the theory of inflation. 

An inflationary cosmology consistent with the present observable Universe requires that the energy in the inflaton must have been transferred into matter degrees of freedom to ignite the hot Big Bang in time for Big Bang nucleosynthesis \cite{Yang:1983gn}---the Universe must be reheated. Yet reheating remains one of the most poorly understood epochs of our cosmic history due to the dearth of observational probes. Because reheating is a local process, the information about its dynamics is largely erased as the standard model plasma reaches local thermal equilibrium. Further, the nonlinear gravitational evolution of structure formation washes out information on scales relevant to preheating. Studies of the physics of preheating with direct observational predictions therefore remain acutely important and timely. 

The nonperturbative decay of the inflaton, known as {\sl preheating} \cite{Traschen:1990sw, Shtanov:1994ce,Kofman:1994rk, Kofman:1997yn}, often occurs at the end of inflation; the explosive production of particles during this epoch can lead to distinctive gravitational  signatures that persist through the opaque, hot, dense phases of  early expansion of the Universe. Among these signatures are the production of a high-frequency stochastic gravitational wave background \cite{Khlebnikov:1997di, Easther:2006gt, Easther:2006vd, Garcia-Bellido:2007nns, Easther:2007vj, Dufaux:2007pt, Dufaux:2008dn, Dufaux:2010cf, Adshead:2018doq, Adshead:2019lbr, Adshead:2019igv, Cosme:2022htl},  and  collapsed compact objects such as primordial black holes \cite{Bassett:1998wg, Green:2000he, Jedamzik:2010dq, Martin:2019nuw, Musoke:2019ima, Auclair:2020csm, Eggemeier:2021smj} or compact minihalos ~\cite{Bringmann:2011ut, Aslanyan:2015hmi}. Non-gravitational signatures include the production of primordial magnetic fields \cite{Adshead:2016iae}, and the possible generation of the baryon asymmetry, \cite{Anber:2015yca, Adshead:2015jza, Kamada:2016eeb, Caldwell:2017chz, Adshead:2017znw, Domcke:2019mnd, Domcke:2022kfs}. These signatures may provide important observational evidence of the reheating and post-inflationary epochs and lead to clues about the microphysics of the early Universe. 

In this paper, motivated by observationally favored, non-minimal classes of inflationary model involving kinetically coupled scalar fields \cite{Linde:2018hmx, Braglia:2020eai, Kallosh:2022vha}, we study a type of kinetic preheating in which the inflaton is coupled to a second scalar field via a dilaton-like interaction \cite{Linde:2018hmx, Kallosh:2022feu, Adshead:2023nhk}. We specialize to exponential-type couplings that arise in dilaton-axion theories, where the dilaton drives inflation. We restrict our attention  to scenarios where the axion does not play a role during inflation, and enters the reheating phase with no vacuum expectation value (VEV). Generalizing our previous work \cite{Adshead:2023nhk}, we explore the effects of different potentials on preheating and study the production of gravitational waves. We characterize the conditions under which the predicted gravitational wave spectra can be constrained by present or next-generation CMB measurements of the effective number of relativistic degrees of freedom, $N_{\rm eff}$. Importantly, the class of models we study here can generically produce gravitational wave backgrounds loud enough that CMB bounds on $N_{\mathrm{eff}}$ may provide constraints on these models. 

This paper is organized as follows. In \cref{sec:model} we describe the model and derive the equation of motion for the fields and the background FLRW spacetime. We then analyze the growth of small fluctuations during the coherent oscillations at the end of inflation and validate our code using a Floquet analysis in \cref{subsec:linanaly}. In \cref{sec:numerics} we describe the numerical methods we use to study the preheating period, and in \cref{sec:results} we discuss our results, in particular characterizing the reheating efficiency and gravitational wave production in the models we study. Our conclusions are presented in \cref{sec:conclusions}.

We use natural units, $\hbar = c = 1$, and define the reduced Planck mass $\Mpl = 1/\sqrt{8\pi G}$. We use the ``mostly plus"  metric convention and repeated/contracted Greek spacetime indices are summed via the Einstein summation convention.

\section{The Model \label{sec:model}}

We consider a theory with an inflaton, $\phi$, and a second scalar field, $\chi$, minimally coupled to Einstein gravity described by the Langrangian
\begin{align}\label{action}
\mathcal{L} = -\frac{M_{\rm Pl}^2}{2}R -\frac{1}{2}\left(\partial\phi\right)^{2} - \frac{W(\phi)}{2}(\partial\chi)^2 - V(\phi) - \frac{m^{2}_{\chi}}{2}\chi^2,
\end{align}
where $R$ is the Ricci scalar.  The inflaton interacts with $\chi$ via an exponential dilaton-like coupling  \cite{Braglia:2020eai, Kallosh:2022vha},
\begin{equation}
W(\phi) = e^{2\phi/\mu}.    \label{coupling term}
\end{equation}
For this work, we assume that the field $\chi$ has no VEV during inflation. Note that this configuration is stable, as shown in Ref.\ \cite{Achucarro:2017ing}.

During inflation, potentials that {\sl plateau}, e.g. polynomial attractors \cite{Kallosh:2022feu} or $\alpha$-attractors \cite{Linde:2018hmx}, are favored by the CMB~\cite{Planck:2018jri,BICEP:2021xfz}.  Phenomenologically, $\alpha$-attractors are particularly interesting as, once the normalization of the scalar power spectrum is specified, the single parameter $\mu$ that controls the remaining freedom in specifying the potential also controls the kinetic coupling.  We consider potentials in this class including the asymmetric E-model $\alpha$-attractor \cite{Linde:2018hmx}
\begin{equation}
V = \frac{m^2\mu^2}{2}\left(1-e^{-\frac{\phi}{\mu}}\right)^2,
\label{eq:Emodel}
\end{equation}
the symmetric T-model $\alpha$-attractor 
\begin{equation}
V = \frac{m^2\mu^2}{2}\tanh^{2}\left(\frac{\phi}{\mu}\right),
\label{eq:Tmodel}
\end{equation}
and the polynomial $\alpha$-attractor \cite{Kallosh:2022feu}
\begin{align}
V = \frac{m^2\mu^2}{2}\frac{\phi^2}{\phi^2+\mu^2}.
\label{eq:Pmodel}
\end{align}
Depending on the choice of $\mu$, these potentials lead to different phenomenology at the end of inflation.  In the limit where $\mu\gg M_{\rm pl}$, the potential is very well approximated as a quadratic.
For smaller values of $\mu\lesssim M_{\rm pl}$, such as those studied in \cite{Aurrekoetxea:2023jwd}, the potential causes  anharmonic oscillations of the $\phi$ field.  These anharmonic contributions during preheating generate self-resonances that cause the homogeneous mode of $\phi$ to decay.  In the absence of a coupled field, these anharmonic oscillations have been shown to create oscillons \cite{Bogolyubsky:1976nx,Bogolyubsky:1976sc,Gleiser:1993pt,Copeland:1995fq,Kasuya:2002zs,Saffin:2006yk,Hertzberg:2010yz,Amin:2011hj,Salmi:2012ta,Antusch:2017flz,Hasegawa:2017iay,Gleiser:2019rvw,Antusch:2019qrr,Ibe:2019vyo,Zhang:2020bec,vanDissel:2023zva} and produce gravitational waves \cite{Antusch:2016con,Amin:2018xfe}. While these self-resonances also occur in the model presented here, we demonstrate that the additional instabilities in the axion field can dominate the preheating phase.

The action, \cref{action}, leads to the classical equations of motion for the fields $\phi$ and $\chi$,
\begin{align}
    \ddot{\phi} &= - 3\frac{\dot{a}}{a}\dot{\phi} + \frac{\nabla^2 \phi}{a^2} -\frac{\partial V}{\partial \phi}+ \frac{e^{2\phi/\mu}}{\mu}\left(\dot{\chi}^2 -\frac{\left(\nabla\chi\right)^2}{a^2}\right) ,
   \label{eq:eomphi} \\
   \nonumber
\ddot{\chi} &= - 3\frac{\dot{a}}{a}\dot{\chi} + \frac{\nabla^2 \chi}{a^2} -\frac{1}{e^{2\phi/\mu}}m^2_\chi \chi \\ & \qquad\qquad - \frac{2}{\mu}\left(\dot{\chi}\dot{\phi} -\frac{\left(\vec{\nabla}\chi\right)\cdot\left(\vec{\nabla}\phi\right)}{a^2} \right).
\label{eq:eomchi}
\end{align}
Here and in what follows, an overdot represents a derivative with respect to cosmic time. 
The expansion of the Universe is given by the Friedmann constraint,
\begin{equation}
H^2 = \frac{1}{3\Mpl^2}\rho.
\label{eq:friedconst}
\end{equation}
Here $H = \dot{a}/a$, and $\rho = \rho_\phi+\rho_\chi$, where
\begin{equation}
\rho_\phi = \dot{\phi}^2 + \frac{\left(\nabla\phi\right)^2}{a^2} + \phi^2,
\label{eq:fullrho}
\end{equation}
and
\begin{equation}
\rho_\chi = e^{\frac{2\phi}{\mu}}\left(\dot{\chi}^2 + \frac{\left(\nabla\chi\right)^2}{a^2}\right) + m_{\chi}^{2}\chi^2. 
\end{equation}

The  energy density of the $\chi$ field relative to the total energy density, $\rho_\chi / \rho$, serves as a useful figure of merit to characterize the efficiency of preheating; complete reheating occurs when this ratio approaches unity.

In all three $\alpha$-attractor models, \cref{eq:Emodel,eq:Tmodel,eq:Pmodel}, we are free to chose one parameter, in our case $\mu$. Planck \cite{Planck:2018jri} constrains the ratio of the squared Hubble rate, $H^2$, and the slow roll parameter, $\epsilon = -\dot{H}/H^2$, via the amplitude of scalar fluctuations,
\begin{equation}
\frac{H_{\rm 50}^2}{8\pi \Mpl^2 \epsilon_{\rm 50}} = 2\times 10^{-9},
\label{eq:setmfromepsilon}
\end{equation}
which fixes $m_\phi$.  Here, $H_{\rm 50}$ and $\epsilon_{\rm 50}$ are evaluated 50 $e$-foldings before the end of inflation; these quantities as well as $m_\phi$ vary from one value of $\mu$ (and one choice of $V(\phi)$) to another.

\section{Stability analysis}\label{subsec:linanaly}

\begin{figure*}[t]
\centering
\includegraphics[width=\linewidth]{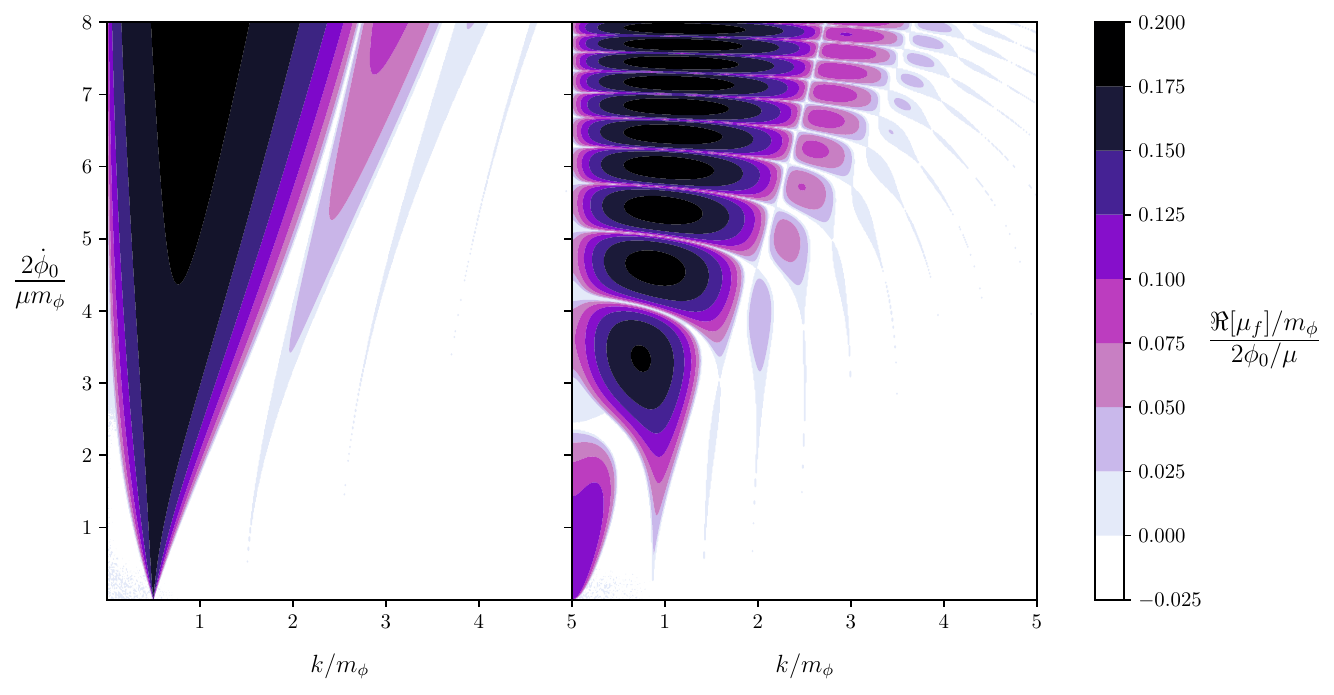}
\caption{\label{fig:floquet} Contours of the real part of the Floquet exponent showing the regions of parameter space where the modes grow exponentially in  the case where $\chi$ is massless (left) and massive (right).   The parameter space corresponding to exponentially growing modes is significantly different in the two cases, and importantly, for a massive $\chi$ field there exists parameter space for exponential growth for modes of arbitrarily long wavelength, even for very small values of $\dot{\phi}_0/\mu m_{\phi}$. The wavenumbers on the horizontal axis correspond to {\sl physical} wavenumbers since this analysis assumes a Minkowski spacetime.}
\end{figure*}
Preheating is defined by the parametric or tachyonic amplification of specific bands of  modes (of the inflaton or other matter fields to which it is coupled), generally sourced by the homogeneously oscillating mode of the inflaton.  We employ Floquet analysis to identify scales on which instabilities exist in these models.  We first linearize the equations to study their stability and predict where the instability bands (in the $\chi$ field) might occur. We then explore the stability of the linearized solutions about the homogeneous oscillating inflaton background and compare them to our simulations.  

While the linearized analysis makes a number of assumptions---both about the self-interactions of the fields and the expansion of the Universe---it is a way to validate our simulations and to gain intuition for the dependence of preheating on the model parameters. In this section, we restrict attention to the quadratic potential, where the inflaton oscillations are harmonic in Minkowski spacetime. 

\subsection{Linearized equations}\label{subsec:linearize}

Ignoring spatial gradients, and neglecting the backreaction of the $\chi$ field, the equation of motion for the homogeneous mode of the inflaton is
\begin{align}\label{eq:decoupledphi}
\ddot{\phi} + 3H\dot{\phi} + m_\phi^2 \phi = 0.
\end{align}
Assuming that the fluctuations of the $\chi$ field are linear perturbations, we can study their independent Fourier modes, 
\begin{align}
\chi(t,{\bf x}) = \int \frac{d^3 k}{(2\pi)^3} \chi_{\bf k}(t)e^{i{\bf k}\cdot{\bf x}},
\end{align}
which are subject to the linearized equation of motion,
\begin{align}
\ddot{\chi}_{\bf k} + \left(3\frac{\dot{a}}{a}- \frac{2}{\mu}\dot{\phi}\right) \dot{\chi}_{\bf k} + \left(\frac{k^2 }{a^2} +\frac{m^2_\chi}{e^{2\phi/\mu}} \right)\chi_{\bf k} = 0.\label{eq:chilineareom}
\end{align}
We rescale $\chi$ to bring its kinetic term to canonical form,
\begin{align}
 \varphi_{\bf k} \equiv z \chi_{\bf k}  =  a^{3/2} \sqrt{{W(\phi)}}\chi_{\bf k},
 \label{eq:varphieom}
\end{align}
so that  \cref{eq:chilineareom} becomes
\begin{align}\label{eq:chilinear}
\ddot{\varphi}_{\bf k} + \left(\frac{k^2}{a^2}+\frac{m^2_\chi}{e^{2\phi/\mu}} -\frac{\ddot{z}}{z}\right)\varphi_{\bf k} = & 0.
\end{align}
Substituting $W(\phi) = e^{2\phi/\mu}$, we can evaluate
\begin{align}
\frac{\ddot{z}}{z} = \left(\frac{3}{2} H +\frac{ \dot{\phi}}{\mu}\right)^2+ \left(\frac{3}{2} \dot{H} +\frac{ \ddot{\phi}}{\mu} \right).
\end{align}
The analogous, linearized equation of motion for the modes of the canonically-normalized inflaton, $\pi_{\bf k} \equiv a^{3/2}\phi_{\bf k}$, reads
\begin{equation}
\ddot{\pi}_{\bf k} + \left(\frac{k^2}{a^2}+\frac{\partial^2 V}{\partial \phi^2} -  \frac{9}{4} H^2 - \frac{3}{2} \dot{H}  \right)\pi_{\bf k}  =  0.
\label{eq:philinear}
\end{equation}
We use \cref{eq:philinear} to set initial conditions for the fluctuations of the inflaton in  our simulations.

\subsection{Floquet analysis}\label{subsec:floquet}

\begin{figure*}
\centering
\includegraphics[width=\linewidth]{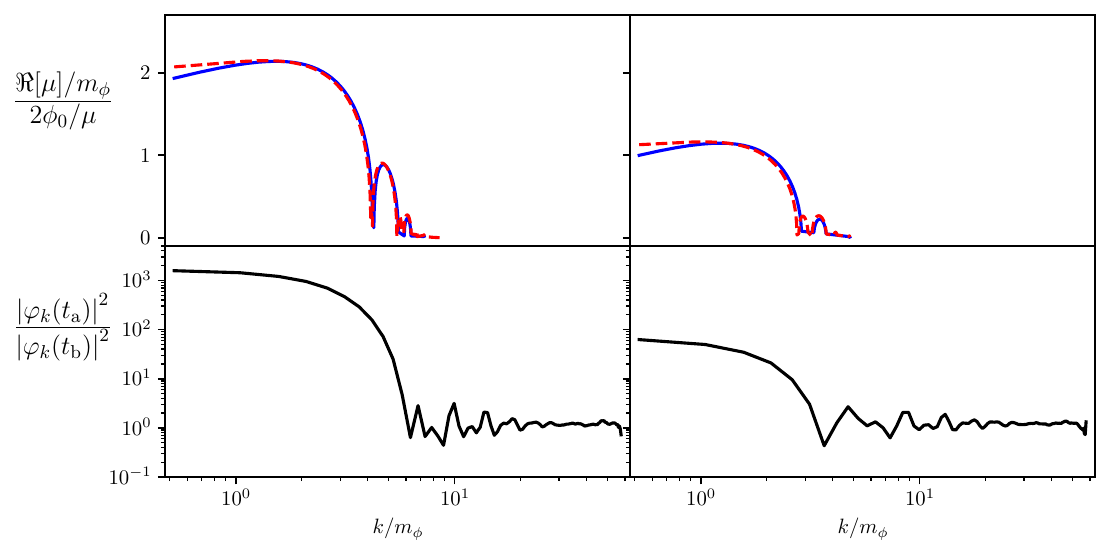}
\caption{\label{fig:floquettwocase} A comparison of Floquet analysis and mode amplification. The top panels show the Floquet exponent, $\mu$, as calculated from \cref{eq:chilinear} (red, dashed) and \cref{eq:chilinearmink} (blue, solid); curves terminate when the Floquet exponent is identically zero.  The bottom panel shows the amplification of the power spectrum of our simulations, $\left|\varphi_k\right|^2$, evaluated just before the first zero crossing of the homogeneous mode of $\phi$, $t_{\rm b}$, and just after, $t_{\rm a}$. The time interval is symmetric about the zero crossing with width $t_{\rm a} - t_{\rm b} \approx m_\phi^{-1}$.  We compare these for two different scenarios: the left panels show the case where $\mu \approx 0.122\, \Mpl$ (corresponding to $2\dot{\phi}/\mu m_{\phi} \approx 7.92$ during the first zero crossing) and the right panel shows the case where $\mu \approx 0.217 \, \Mpl$ (corresponding to $2\dot{\phi}/\mu m_{\phi} \approx 4.46$ during the first zero crossing).  Note that the relative amplitude and the range of $k$-modes that are amplified show agreement with \cref{fig:floquet}. }
\end{figure*}

We begin by studying the behavior of eqs.\ \eqref{eq:decoupledphi} and \eqref{eq:chilinear} in Minkowski space, which can be done analytically. In this limit, $a = 1$ and $H = \dot{H} = 0$, and hence the solution for the homogeneous inflation field is 
\begin{align}
\phi = \phi_0\cos(m_\phi t).
\end{align}
Under these assumptions, the equation of motion for the modes of $\varphi$ is
\begin{align}\label{eq:chilinearmink}
\ddot{\varphi}_{\bf k} + \left(\frac{k^2}{a^2}+m_\chi^2 e^{-2\phi/\mu} -  \left(\frac{ \dot{\phi}}{\mu}\right)^2-\frac{ \ddot{\phi}}{\mu}  \right)\varphi_{\bf k}   =  0,
\end{align}
which is the equation for a harmonic oscillator with a time-dependent (periodic) effective mass. 
Floquet's theorem states that solutions to \cref{eq:chilinearmink} are of the form
$\varphi_{\bf k} \propto P_{+}(t)e^{\mu_f t} + P_{-}(t)e^{-\mu_f t}$, where $\mu_f$ is the {\sl Floquet exponent}
and $P_\pm (t) = P_\pm(t + T )$ is a periodic function with period $T$ which is the same period as the time-dependent mass. When $\Re(\mu)\neq 0$, the modes, $\varphi_{\bf k}$, grow exponentially. 

In the small amplitude limit, $\phi_0 \ll \mu$, \cref{eq:chilinearmink} can be recast as the Mathieu equation, 
\begin{align}\label{eq:mathieu}
\frac{d^2 \varphi_k}{dz^2} + [p-2q\cos(2z)]\varphi_k = 0,
\end{align}
where $z = m_\phi/2$ and we have defined
\begin{align}
p = & \left(\frac{2k}{m_\phi}\right)^2 +\left(\frac{2m_{\chi}}{m_{\phi}}\right)^2,\\
q =& \frac{2\phi_0}{M}\frac{m_\chi^2}{m_\phi^2}\left(1-\frac{m_{\phi}^2}{2m_\chi^2}\right).
\end{align}
When $\phi_0\ll \mu$, $q \ll 1$, the solutions of the Mathieu equation are known to be unstable when $ p = n^2$ for $n\in \mathbf{Z}$. Notice that, for $k\ll m_\phi$, if $m_\chi = n m_\phi/2$, then the instability band persists down to arbitrarily small wavenumbers; this can also be seen in the right panel of \cref{fig:floquet} when directly using \cref{eq:chilinearmink}. This effect is analogous to that which occurs for kinetically coupled massive vector fields \cite{Adshead:2023qiw}.

Floquet exponents can be obtained analytically for small $q$, and for $m_\chi =  m_\phi / 2$ they are given by 
\begin{align}
\Re[\mu] = \frac{\phi_0}{\mu}\frac{m_\phi}{4}.
\end{align}
Away from this limit, for both large $q$ and $m_\chi \neq  m_\phi/2$, the Floquet exponents must be found numerically. These are displayed in \cref{fig:floquet}, which shows the Floquet exponents, $\mu_f$, as the amplitude of the velocity of the homogeneous mode, $\dot{\phi}_0 = m_\phi \phi_0$, is varied relative to the kinetic coupling, $\mu$, for a set of wavenumbers using \cref{eq:chilinearmink}.  

\section{Numerical Procedure}\label{sec:numerics}

We employ {\sc GABE} \cite{Child:2013ria}  to carry out 3+1 simulations of the fields $\phi$ and $\chi$ in a self-consistent, but rigidly expanding background (we ignore local gravity).  To set our initial conditions, we evolve the  equations of motion in the homogeneous limit, \cref{eq:eomphi} alongside the Friedmann constraint, \cref{eq:friedconst} for at least 60 $e$-folds of inflation. This ensures that the system is following the corresponding attractor solution for each model we study. From this homogeneous evolution, we extract $\epsilon_{50}$ and $H_{50}$ which allow us to calculate $m_\phi$ for every choice of potential and $\mu$.  These also give the inflaton amplitude and velocity half an $e$-folding before the end of inflation (inflation ends, by definition, when $\ddot{a} = 0$).  We initialize the fluctuations of the inflaton and $\chi$ fields from the Bunch-Davies vacuum, 
\begin{align}
\langle  |\phi(k)|^2  \rangle &= (2 a \omega_{\phi})^{-1}, \\
  \langle |\chi(k)|^2 \rangle  &= (2 a \omega_{\chi} W(\bar{\phi})))^{-1},
\end{align}
where for each field $\omega = \sqrt{(k/a)^2 +m^2_{\mathrm{eff}} }$ and the effective mass squared for $\phi$ and $\chi$ are calculated from the equations of motion \cref{eq:philinear} and \cref{eq:varphieom}, evaluated half an $e$-folding before the end of inflation. Note the factor of $W$ in the $\chi$ initial spectrum is a consequence of the fact the kinetic term for $\chi$ carries the dilaton-like coupling $W$.  

We start the runs at half an $e$-folding before the end of inflation to ensure that the effective mass of $\phi$ is positive, and the modes of interest are in the Bunch-Davies vacuum.  We initialize our box to have size $L = H^{-1}_{\rm end}e^{-0.5}$, where $H_{\rm end}$ is the Hubble scale at the end of inflation. This  initialization ensures that our box is the size of the Horizon at the end of inflation and the onset of preheating. These initial conditions seed the fully non-linear evolution of the system \cref{eq:eomphi}, \cref{eq:eomchi} alongside the homogenous expansion of the Universe, \cref{eq:friedconst}, from half an e-folding before the end of inflation through the reheating epoch.  Unless otherwise noted, we use grids with $N^3 = 256^3$ points with periodic boundary conditions.  We set the timestep is set to be $\Delta t = L/N/30 < L/N/\sqrt{3}$ to satisfy the Courant–Friedrichs–Lewy condition.

\subsection{Instability in a toy model}\label{sec:toymodel}

Before we turn to the $\alpha$-attractor models, we validate our simulations using the quadratic potential,
\begin{equation}
    V(\phi) = \frac{1}{2} m^2\phi^2,
    \label{eq:toypot}
\end{equation}
where we set $m_{\phi}=5\times 10^{-5} \Mpl$ as a reference value, and compare our full nonlinear results to the linearized Floquet analysis from \cref{subsec:linanaly} above.  In practice $m_\phi$ sets the scale of the simulation and is only relevant for setting the initial conditions of the fields.  In Minkowski space,  a quadratic potential  leads to a harmonic oscillation of the homogeneous value of $\phi$.  While  this is not true in an expanding Universe, when the $\phi$ field crosses zero---the point at which particle creation is most violent---we can instantaneously approximate the magnitude of the sinusoidal oscillation to the velocity of the field at the zero crossing, $\phi_0 = \dot{\phi}_0/m_\phi$. Then we can use this value to predict where the instability bands are.  We perform a set of simulations where $m_\chi = 0$ and $m_\chi = m_\phi/2$ for a range of $\mu$. 

For these runs we are able to start at the end of inflation with an initial comoving box size of $L = 12 \,m_\phi^{-1}\approx 6H_{\rm end}^{-1}$ since the quadratic potential prevents tachyonic modes at the end of inflation.  In this toy model, we can also take coarser resolution, where $N^3 = 128^3$, which roughly probes modes between $k_{\rm min} = 2\pi/L \approx 0.5 m_\phi$ and the 3D Nyquist frequency,  $k_{\rm max} = (128 \sqrt{3}/2) k_{\rm min} = 58 m_\phi$.  \Cref{fig:floquettwocase} shows the comparison between the Floquet analysis for the first zero-crossing of the homogeneous mode of $\phi$ and the results of two fully nonlinear simulations, one with $\mu = 0.122\Mpl$ and one with $\mu = 0.217\Mpl$, both in the case where $m_\chi = 0$.  We see excellent agreement with regards to which modes are amplified and the relative strength of these amplifications.  Note that the results of the nonlinear simulations do not show all the bumps of the Floquet analysis---this is due to the fact that the modes from the nonlinear simulations are binned and that, even over the short timescale over which we compare, the physical size of the modes vary enough to smear out the finest features of the Floquet analysis.  Nonetheless, this agreement validates that kinetic preheating is responsible for the particle creation in these models.

\begin{figure*}
\centering
\includegraphics[width=\linewidth]{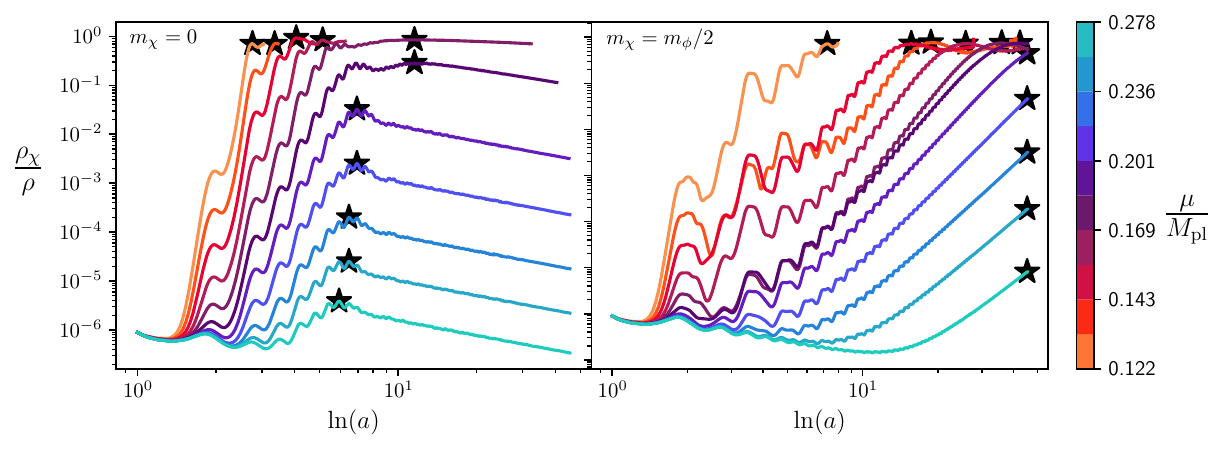}
\caption{\label{fig:ratiosforM}  The reheating efficiency for the toy-model given by \cref{eq:toypot} for a range of $\mu$ values as a function of $e$-folding in the case where $\chi$ is massless (left panel) and massive, $m_\chi = m_\phi/2$ (right panel). Smaller $\mu$ values correspond to a larger dilaton-like couplings $W$, yielding more efficient reheating. Warm colors (cool colors) correspond to the data produced by simulations with a smaller (larger) $\mu$ values, as indicated by the colorbar. The stars represent the points in each run where the ratio $\rho_\chi/\rho$ is largest.  Reheating into the $\chi$ field is faster and more complete for smaller $\mu$ values. }
\end{figure*}

We also calculate the efficiency of preheating in this toy model.  \Cref{fig:ratiosforM} shows the ratio of energy in the $\chi$ field compared to the total energy in the box as a function of scale factor throughout the simulation for a range of values of $\mu$ for both $m_\chi =0$ and $m_\chi = m_\phi/2$.  While we terminate all the runs when the scale factor hits $a\sim 500$, we see that when $\chi$ is massive, the low-$k$ instability continues to source particle creation until we end the simulation. In \cref{app:toyGWs}, we compute the spectrum of gravitational waves from these toy scenarios.

\section{Results}\label{sec:results}

In this section, we present the results of  our nonlinear simulations. We demonstrate that kinetic preheating can be extremely efficient in  $\alpha$-attractor models, across all variants of the potential we study. We also compute the resulting spectrum of gravitational waves, and demonstrate that current and upcoming CMB measurements will constrain kinetic preheating in these models. 

\subsection{Preheating \label{subsec:preheat}}

\begin{figure*}
\centering
\includegraphics[width=\linewidth]{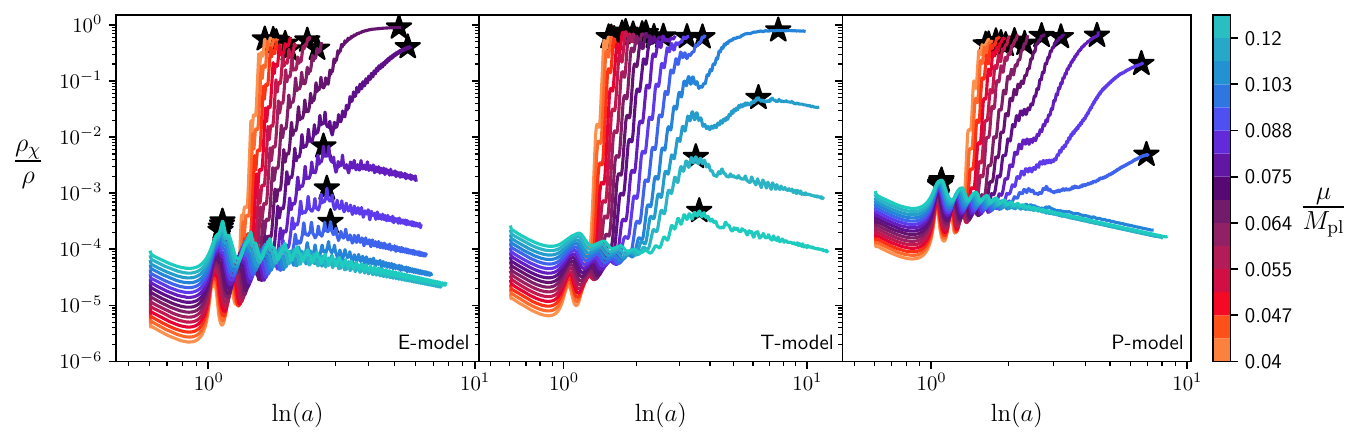}
\caption{\label{fig:EPTratios} Reheating efficiency, $\rho_\chi/\rho$, for the E-model (left), T-model (center) and P-model (right) $\alpha$-attractor given by \cref{eq:Emodel}, \cref{eq:Tmodel} and \cref{eq:Pmodel} for a range of $\mu$ values as a function of $e$-folding. Smaller $\mu$ values correspond to a larger dilaton-like couplings $W$ (and a broader potential $V(\phi)$), yielding more efficient reheating. Warm colors (cool colors) correspond to the data produced by simulations with a smaller (larger) $\mu$ values, as indicated by the colorbar. The stars represent the points in each run where the ratio $\rho_\chi/\rho$ is largest.}
\end{figure*}

\begin{figure}
\centering
\includegraphics[width=\linewidth]{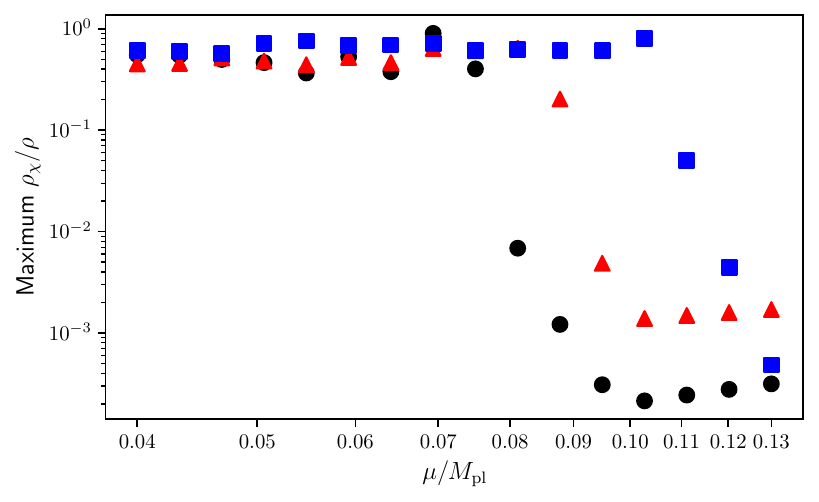}
\caption{\label{fig:ratiosvsm} Comparison of maximum ratio of energy densities $\rho_\chi/\rho$ between the three different $\alpha$-attractor models, the E-model (black, circles), T-model (blue, squares), and P-model (red, triangles) over a range of $\mu$ values, demonstrating the critical $\mu$ value for which reheating becomes efficient in each model. These are the maxima that correspond to the starred points in \cref{fig:EPTratios}.  The E-model generally requires the smallest $\mu$ value to reheat efficiently; by contrast, the T-model can reheat efficiently at larger $\mu$ values. 
}
\end{figure}

We begin by examining the reheating efficiency in the three $\alpha$-attractor models from \cref{sec:model}.  \Cref{fig:EPTratios} shows the ratio $\rho_{\chi}/\rho$ as a function of efolding number for a range of $\mu$ values for the E-model, P-model, and T-model in the left, center and right panels, respectively. In all models preheating is more efficient as $\mu$ is decreased. For a fixed $\mu$ value, we find that reheating is most (least) efficient for the T-model (E-model).   This can be seen more clearly in \cref{fig:ratiosvsm}, where we show the maximum energy density ratio $\rho_{\chi}/ \rho$ achieved in each simulation as a function of $\mu$. Generally, for each model there is a sharp transition between inefficient ($\rho_\chi/\rho \ll 1$) reheating and complete reheating $\rho_\chi/\rho \sim 1$ that occurs when $\mu$ drops below a critical value. This transition occurs at larger $\mu$ values for the T-model and lower values for the E-model, and beyond this transition all three models reach roughly the same maximum value of $\rho_{\chi}/\rho$. Finally, for each model, our results demonstrate that if $\mu$ is small enough, preheating proceeds very rapidly and violently, causing the simulations to crash before reheating can fully complete (see the cool colored curves that end abruptly in  \cref{fig:EPTratios}). This is a physical consequence of the fact that  the kinetic coupling $W$ generates an infinite sequence of higher dimensional operators. These operators generate a high-frequency, ultra-violet cascade that quickly transfers power to high frequency modes that our simulations cannot resolve \cite{Deskins:2013dwa,Adshead:2017xll}. 

\subsection{Gravitational Waves}\label{sec:gravwavs}

Strong resonance dynamics during preheating generically leads to the robust generation of gravitational waves \cite{Khlebnikov:1997di, Easther:2006gt, Easther:2006vd, Garcia-Bellido:2007nns, Easther:2007vj, Dufaux:2007pt, Dufaux:2008dn, Dufaux:2010cf, Adshead:2018doq, Adshead:2019lbr, Adshead:2019igv, Cosme:2022htl}. The $\alpha$-attractor models studied here can lead to very efficient kinetic preheating and we anticipate that they may also lead to a strong gravitational wave background. Here, we calculate the power in gravitational waves by passively calculating tensor perturbations of the metric, following the basic procedure of ~\cite{Easther:2007vj}.  Ignoring scalar and vector perturbations, the metric reads
\begin{equation}
{\rm d}s^2 = -{\rm d}t^2 + a^2\left(\delta_{ij} + h^{TT}_{ij}\right){\rm d}x^i {\rm d}x^j.
\end{equation}
 Here, the transverse-traceless perturbations of the spatial metric, $h^{TT}_{ij}$, are gravitational waves. These satisfy the linearized equation of motion
\begin{align}\label{eqn:GWeom}
\Box h^{\rm TT}_{ij} = 16 \pi G \,T^{\rm TT}_{ij},
\end{align}
which follows from the Einstein equation. Gravitational waves are sourced by the transverse-traceless projection of the anisotropic stress tensor,
\begin{equation}
    T_{ij}^{\rm TT} = \left(P_{il} P_{jm} - \frac{1}{2} P_{ij} P_{lm} \right) T_{lm},
\end{equation}
where $P$ is the projection operator,
\begin{equation}
	P_{ij} = \delta_{ij} - \frac{k_i k_j}{k^2}.
\end{equation}
\Cref{eqn:GWeom} allows us to compute the evolution of $h_{ij}$, from which we can compute the effective stress energy tensor for gravitational waves ~\cite{Misner:1973prb},
\begin{equation}
    T_{\mu \nu}^{\rm gw} = 8\pi G \left<h_{ij,\mu}^{\rm TT}{{h^{ij}}_{,\nu}}^{\rm TT}\right>.
\end{equation}
The energy density in gravitational waves is therefore
\begin{align}
\rho_{\rm gw} = 8\pi G \left|h_{ij,0}^{\rm TT}\right|^2,
\end{align}
and the spectral energy density of the gravitational waves during the simulation is found from 
\begin{equation}
\label{omegagw}
    \Omega_{\rm gw}(k) 
    \equiv \frac{1}{\rho} \frac{d \rho_\mathrm{gw}}{d \ln k} 
	= \frac{1}{24\pi^2 L^3} \frac{k^3}{H^2} \sum_{i, j} \left\vert \dot{h}_{ij}(k, t) \right\vert^2.
\end{equation} 
To compare between simulations, we evaluate the gravitational wave spectra when $\rho_{\chi}/\rho$ is at its maximum. This occurs at the points identified with stars in \cref{fig:EPTratios}.  Once computed, the spectral energy density in gravitational waves today can be found from \cite{Easther:2006gt,Easther:2006vd}
\begin{align}\label{eqn:GWtransfer}
    \Omega_{\rm gw, 0} h^2
    &= \Omega_{\rm rad, 0} h^2
        \frac{g_{\star}(a_\mathrm{r})}{g_{\star}(a_0)}
        \left( \frac{g_{\star S}(a_\mathrm{r})}{g_{\star S}(a_0)} \right)^{-4/3}
        \Omega_\mathrm{gw}(a),
\end{align}
with frequencies today given by 
\begin{align}
    f
    &= \frac{k / 2 \pi a}{\rho(a)^{1/4}}
        \rho_{\mathrm{rad}}(a_0)^{1/4}
        \left( \frac{g_{\star}(a_\mathrm{r})}{g_{\star}(a_0)} \right)^{1/4}
        \left( \frac{g_{\star S}(a_\mathrm{r})}{g_{\star S}(a_0)} \right)^{-1/3}
    \\
    &\approx 3.2 \times 10^{10} \, \mathrm{Hz} \frac{k / a}{\sqrt{H(a) \Mpl}}
        \left( \frac{g_{\star}(a_\mathrm{r}) / g_{\star}(a_0) }{100} \right)^{-1/12} \label{eq:gwfreq}.
\end{align}
In the preceding expressions $g_\star$ is the number of ultra-relativistic degrees of freedom evaluated at reheating, $a_r$, or today, $a_0$, $g_{\star}(a_\mathrm{r})/g_{\star}(a_0)\approx 100$.  We also make the standard assumption that the Universe is radiation-dominated at the time when the power spectrum in the simulation is evaluated.\footnote{Note that, the transfer functions in \cref{eqn:GWtransfer} assume continuous radiation domination from the time of gravitational wave emission until matter-radiation equality at redshift $z \sim 3000$. However, in cases where the equation of state deviates from $w = 1/3$ due to the domination of some exotic species, these transfer functions are modified.} We also note that we are limited in the range of frequencies for which we can compute the gravitational wave spectra, bounded below by the size of the box and bounded above by the Nyquist frequency of the grid. Nevertheless, we are able to resolve the gravitational wave spectra over several decades in frequency around $10^{10} \,\mathrm{Hz}$, as indicated by \cref{eq:gwfreq}.
\begin{figure*}
\centering
\includegraphics[width=\linewidth]{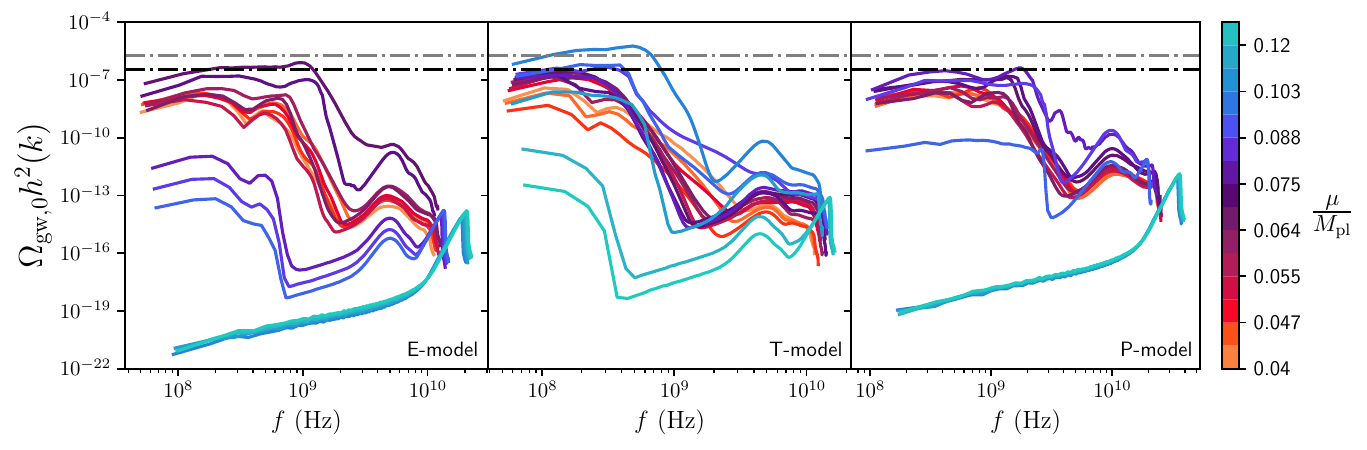}
\caption{\label{fig:EPTmodelGW} The gravitational wave spectral energy density for the E-model (left), T-model (center), and T-model (right)  $\alpha$-attractor kinetic reheating scenarios for a range of $\mu$ values. These spectra are evaluated at the starred-locations in \cref{fig:EPTratios}. Simulations that reheat completely and last longer produce stronger gravitational wave backgrounds that are particularly enhanced at the low frequency end of the resolved spectrum. The dot-dashed lines show the $2\sigma$ constraints on the amplitude of the gravitational wave spectrum that can be inferred from from current (Planck \cite{Planck:2018vyg}, grey) and future (CMB-S4 \cite{Abazajian:2019eic}, black) CMB measurements of $N_{\rm eff}$.  The smaller, high-frequency peaks in these spectra are a numerical artifact.}
\end{figure*}

We present the gravitational wave spectra from preheating in the three $\alpha$-attractor scenarios over a range of $\mu$ values in  \cref{fig:EPTmodelGW}. The spectra in each model share some similarities, but in general the spectra depend non-trivially on $\mu$. Roughly speaking, the gravitational wave spectra are strongest (that is, they reach a largest maximum value at some $f$) for values of $\mu$ near the critical value (the largest value of $\mu$ for which a model preheats). For these values, the physics of reheating is sufficiently violent to produce a strong gravitational wave spectrum, but the process also persists for longer, and the combination of these two factors leads to the strongest gravitational wave spectra \cite{Giblin:2014gra}. 

For models where $\mu$ at and below the critical value necessary for complete reheating, reheating is very fast and the simulations end very quickly as described in \cref{subsec:preheat}, yielding somewhat weaker gravitational wave spectral energy density, with similar spectral shapes.  We note that the ultra-violet cascade in the field sector makes evolving the simulations any further prohibitive in our setup; more sophisticated numerical techniques, larger grids, and longer run times might be able to resolve preheating in these models. The gravitational wave spectra we evaluate in these scenarios are likely lower bounds on the actual generated spectra. 

\begin{figure}
\centering
\includegraphics[width=\linewidth]{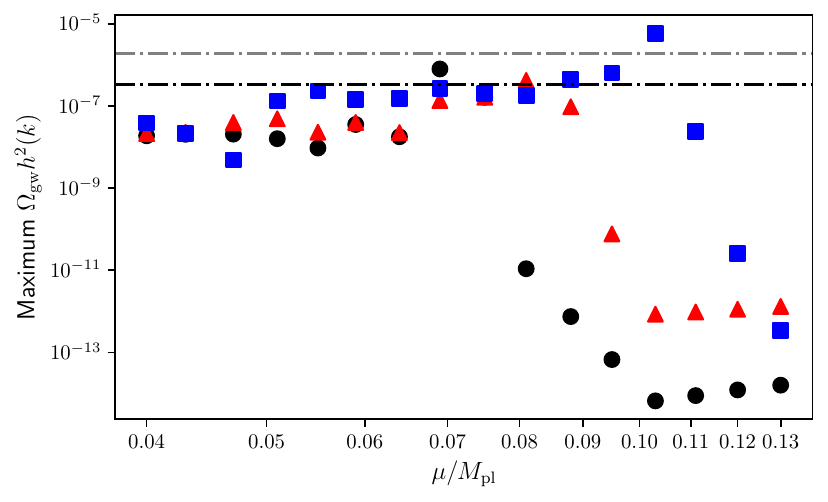}
\caption{\label{fig:maxgwpowerEPT} The maximum gravitational wave spectral energy density from kinetic preheating for each $\alpha$-attractor model, the E-model (black, circles), T-model (blue, squares), and P-model (red, triangles) over a range of $\mu$ values. Each point represents the maximum value of the gravitational wave spectra presented in \cref{fig:EPTmodelGW}. The gravitational wave background is significantly stronger for models that completely reheat. The dot-dashed lines show the $2\sigma$ constraints on the amplitude of the gravitational wave spectrum that can be inferred from from current (Planck \cite{Planck:2018vyg}, grey) and future (CMB-S4 \cite{Abazajian:2019eic}, black) CMB measurements of $N_{\rm eff}$.
}
\end{figure}

Importantly, in all three $\alpha$-attractor models, the gravitational wave spectra can reach $\Omega_{\rm gw,0} \approx 10^{-6} - 10^{-7}$ over a range of scales. A gravitational wave background of this strength contributes significantly to the radiation content of the early Universe and can therefore potentially be constrained by bounds on extra relativistic degrees of freedom, $\Delta N_{\mathrm{eff}}=N_{\mathrm{eff}} - 3.044$ \cite{PhysRevD.93.083522,deSalas:2016ztq, Akita:2020szl}. If we take the conservative simplifying assumption there are no additional ultra-relativistic degrees of freedom at recombination beyond the Standard Model and the gravitational waves from preheating, then the present-day gravitational wave energy density is related to the present day photon energy density and $\Delta N_{\mathrm{eff}}$ via
\begin{align}
    \Omega_{\mathrm{\rm gw,0}} h^2  = \Omega_{\gamma,0}h^2 \frac{7}{8}\left( \frac{4}{11} \right)^{4/3} \Delta N_{\mathrm{eff}},
\end{align}
where the present day photon energy density is $\Omega_{\gamma,0}h^2  \approx 2.47 \times 10^{-5}$. Planck 2018 \cite{Planck:2018jri} gives an upper bound of $|\Delta N_{\mathrm{eff}}| < 0.33$ (95\% CL) which yields a constraint on the gravitational wave energy density $\Omega_{\mathrm{gw,0}} h^2 \lesssim 1.851 \times 10^{-6}$. A joint BBN-CMB analysis \cite{Fields:2019pfx} gives a slightly stronger bound $|\Delta N_{\mathrm{eff}}| <  0.168$ (95\% CL), which tightens the constraint to  $\Omega_{\mathrm{gw,0}} h^2 \lesssim 9.424 \times 10^{-7}$. This means that under the assumptions listed here, and depending on the value of $\mu$, these models can already be constrained at the $2\sigma$ level as a consequence of violent gravitational wave production during the preheating epoch. This picture is even more promising when looking ahead to next generation CMB experiments: the projected sensitivity of CMB-S4  is $|\Delta N_{\mathrm{eff}}| <  0.06$ (95\% CL) \cite{Abazajian:2019eic}, which further tightens the constraint on the gravitational wave spectrum to $\Omega_{\mathrm{gw,0}} h^2 \lesssim 3.366 \times 10^{-7}$. In \cref{fig:maxgwpowerEPT}, we show the maximum energy density in gravitational waves for each $\alpha$-attractor model over a range of $\mu$ values as well as the CMB bound from Planck and projected bound from CMB-S4. This indicates that several $\mu$ values are already in $2\sigma$ tension with bounds on $\Delta N_{\mathrm{eff}}$, emphasizing the role that next-generation CMB experiments will play in constraining the parameter space in these models as a consequence of gravitational waves production during the violent preheating epoch.  

%
%%
%%%%
\section{Discussion and Conclusions}\label{sec:conclusions}
%%%%
%%
%

In this paper we have studied gravitational wave production from kinetic preheating after inflation in $\alpha$-attractor potentials. These non-minimal inflationary models are currently among the leading candidates for the theory of inflation.   In these models, one or more scalar fields are non-minimally coupled to the Ricci scalar. After transforming to the Einstein frame and canonically normalizing the inflaton, these models generically lead to exponential couplings between the inflaton and any other degree of freedom in the theory. Motivated by these models, in this work we have investigated the effect of an exponential kinetic coupling between the inflaton and an ultralight spectator field during the onset of reheating after $\alpha$-attractor inflation.

Beyond their appeal as inflationary candidates, $\alpha$-attractor models also have the attractive feature that they effectively contain only one free parameter. This parameter characterizes both the shape of the inflationary potential as well as the strength of the couplings of the inflaton to other degrees of freedom. After the normalization of the scalar power spectrum is fixed, the remaining free parameter sets the tensor-to-scalar ratio. In this paper we have extended the study of these models to the preheating phase, characterizing the conditions under which these models can efficiently preheat the Universe and produce strong gravitational wave backgrounds. 

Our analysis has shown that kinetic preheating in the three classes of $\alpha$-attractor potentials studied here can be very efficient and is highly sensitive to the sole free parameter $\mu$. For normalization of the scalar spectrum fixed to the observed value, complete reheating is achieved for $\mu/M_{\mathrm{pl}} \lesssim 0.077$ or the E-model, \cref{eq:Emodel}, $\mu/M_{\mathrm{pl}} \lesssim 0.11$ for the T-model, \cref{eq:Tmodel}, and $\mu/M_{\mathrm{pl}} \lesssim 0.81$ for the P-model, \cref{eq:Pmodel}. Because of the exponential sensitivity of the kinetic coupling to $\mu$, even slightly smaller values of $\mu$ yield significantly faster preheating. That is, there exists an upper bound on $\mu$ below which both the inflationary predictions agree with the CMB and the kinetic preheating is highly efficient. 

We have shown that  highly efficient preheating in these models leads to the creation of a gravitational wave background from the preheating phase, with an amplitude and spectral energy density that far exceeds the stochastic inflationary background in the same frequency band. This frequency band is well above the range probed by both current and next generation direct detectors such as the Advanced Laser Interferometer Gravitational-Wave Observatory \cite{LIGOScientific:2014pky}, the Laser Interferometer Space Antenna \cite{LISA:2017pwj}, Cosmic Explorer \cite{LIGOScientific:2016wof}, and the Einstein Telescope \cite{Hild:2010id}. However, we have demonstrated that in some cases, these backgrounds are strong enough to be constrained at the  $2\sigma$ level by present and next-generation CMB bounds on $N_{\mathrm{eff}}$. Kinetic-preheating in $\alpha$-attractor scenarios is therefore both a highly predictive and also falsifiable theory of the primordial Universe, which is desirable given the challenge of constructing concrete early Universe models that connect with present or near-term observations.

The present work suggests several promising avenues for further investigation into kinetic preheating. For example, while our analysis has demonstrated intense gravitational wave production, further work is needed to understand the full gravitational dynamics in these scenarios. In particular, we plan to investigate the fully nonlinear gravitational dynamics by implementing full general relativity in the simulations to study the potential for the production of primordial black holes and other compact objects \cite{Giblin:2019nuv, Adshead:2023mvt,Adshead:2024_2}.

\acknowledgements
We thank Mustafa Amin and Zachary Weiner for early conversations regarding kinetic couplings. We thank Reid Pfaltzgraff-Carlson for early collaboration on this project. P.A.\ and A.J.T.\ are supported in part by the United States Department of Energy, DE-SC0015655.
J.T.G.\ is supported  in part by the National Science Foundation, PHY-2309919.  The authors gratefully acknowledge support from the Simons Center for Geometry and Physics, Stony Brook University at which some of the research for this paper was performed.  P.A. and J.T.G. thank the Yukawa Institute for Theoretical Physics at Kyoto University, where some of this work was completed during the YITP-T-23-02 on ``Cosmology and Gravity 2024." Simulations were performed using the Ohio supercomputer center \cite{OhioSupercomputerCenter1987} as well as on hardware provided by the National Science Foundation, Kenyon College. 

\begin{figure}
\centering
\includegraphics[width=\linewidth]{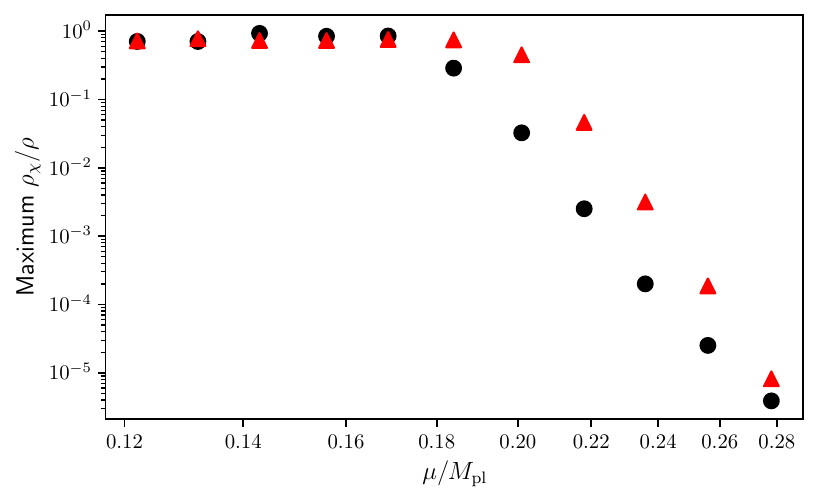}
\caption{\label{fig:quadeff} The preheating efficiency from kinetic preheating with a quadratic potential in the case where $\chi$ is massless (black, circles) and massive, $m_\chi = m_\phi/2$, (red triangles). Massive preheating is more efficient here at higher coupling due to the low amplitude resonance identified in the Floquet analysis in \cref{sec:toymodel} These are the maxima that correspond to the starred points in \cref{fig:EPTratios}.}
\end{figure}

\appendix
\section{Gravitational Waves from a Toy Model}\label{app:toyGWs}

For completeness, in this appendix we compute the gravitational wave spectra from the toy model in \cref{sec:toymodel} where the inflationary potential is quadratic.  In this model, the presence of a non-zero mass in the $\chi$ field only slightly changes the value of the kinetic coupling, $\mu$, required to fully preheat.  \Cref{fig:quadeff} shows that, independent of our two choices for $m_\chi$, $\mu \lesssim 0.18 M_{\rm pl}$ yields efficient preheating.  At the same time, the low amplitude parametric resonance in the case where $m_\chi = m_\phi/2$ allows for efficient preheating at larger values of $\mu$. In this case, the low-$k$ instabilities source $\chi$-particle production over many $e$-foldings (see \cref{fig:ratiosforM}), and potentially make it possible for the Universe to fully reheat at much larger values of $\mu$.

\begin{figure*}
\centering
\includegraphics[width=\linewidth]{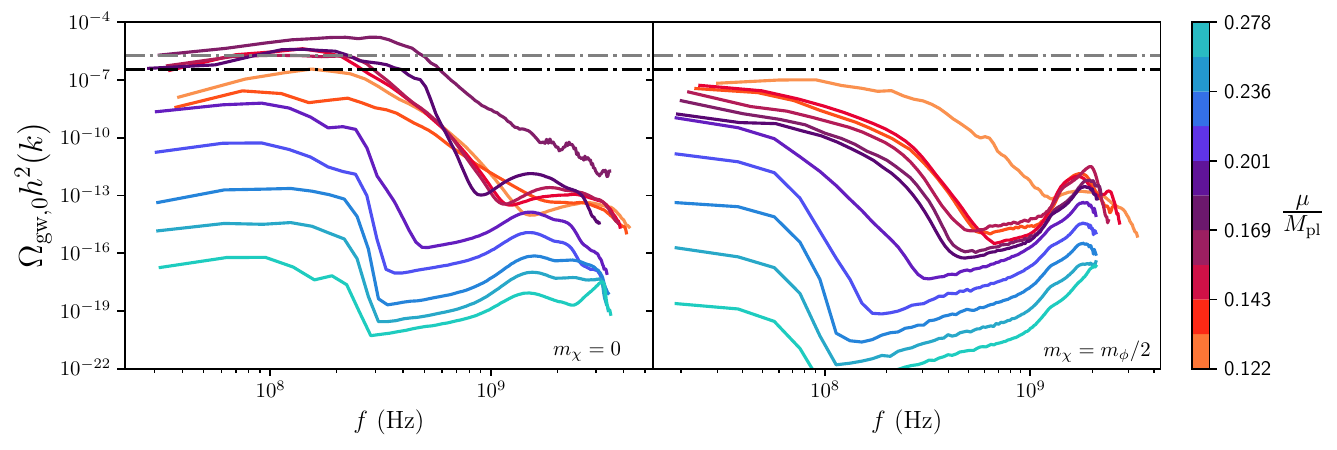}
\caption{\label{fig:mass_nomass_modelGW} The gravitational wave spectral energy density from quadratic potential over a range of $\mu$ values for a massless $\chi$ field (left) and massive $\chi$ field, $m_{\chi} = m_{\phi}/2$ (right). Simulations that reheat completely and last longer produce stronger gravitational wave backgrounds that are particularly enhanced at the low frequency end of the resolved spectrum. These spectra are evaluated at the starred-locations in \cref{fig:ratiosforM}.}
\end{figure*}

\begin{figure}
\centering
\includegraphics[width=\linewidth]{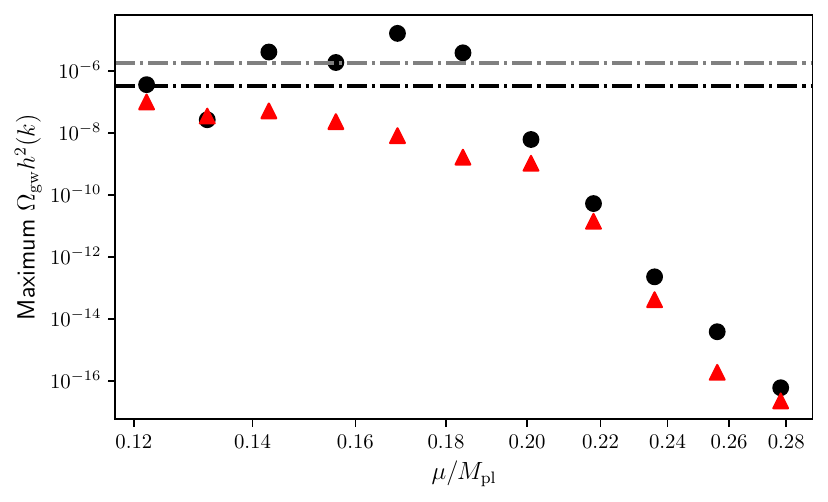}
\caption{\label{fig:maxgwpower_nomass_mass} Maximum spectral energy density in the graviational wave background as a function of coupling $\mu$. The red triangles show the case where $m_\chi = m_\phi/2$, while the massless case, $m_\chi = 0$, is shown in black circles. Each point represents the maximum value of the gravitational wave spectra presented in \cref{fig:mass_nomass_modelGW}.}
\end{figure}

However, we find that the presence of a nonzero mass for the axion, $m_\chi \neq 0$, significantly affects the resulting gravitational wave spectra.  Using the same techniques as described in \cref{sec:gravwavs}, we can compute  the resulting gravitational waves the times identified with stars in \cref{fig:ratiosforM}. These spectra are shown in  \cref{fig:mass_nomass_modelGW}.  Despite the $m_\chi = 0$ and $m_\chi \neq 0$ runs showing similar preheating efficiencies, the amplitude of the gravitational waves produced in the $m_\chi = 0$ case generically exceed those in the $m_\chi \neq 0$, as can be seen in \cref{fig:maxgwpower_nomass_mass}. This effect is pronounced at intermediate values of $\mu$, where the spectra from the massless case exceeds that of the massive case by several orders of magnitude for lower couplings, as we show in \cref{fig:maxgwpower_nomass_mass}. This is particularly relevant for the marginal cases, $\mu \sim 0.18\, M_{\rm pl}$. 

\bibliography{KineticPreheat}

\end{document}